\documentclass[letterpaper, 10 pt, conference]{ieeeconf}  

\IEEEoverridecommandlockouts                              

\usepackage{cite}
\usepackage{amsmath,amssymb,amsfonts}
\usepackage{mathtools}
\usepackage{algorithmic}
\usepackage{graphicx}
\graphicspath{{Figures/}{../Figures/}}
\usepackage{textcomp}
\usepackage{booktabs}
\usepackage{xcolor}
\usepackage{multirow}
\usepackage{cuted}
\usepackage{xstring}
\usepackage{enumerate}
\usepackage{adjustbox}

\providecommand{\com[2]}{\begin{tt}[#1: #2]\end{tt}}
\definecolor{red}{rgb}{1,0,0}

\definecolor{blu}{rgb}{0,0,1}

\definecolor{gre}{rgb}{0,0.7,0.3}

\newcommand{\bydef}{\stackrel{\Delta}{=}}

\title{ Nonlinear VRFT with LASSO}
\author{Alexandre Sanfelici Bazanella, Diego Eckhard
\thanks{This study was financed in part by Conselho Nacional de Desenvolvimento Cient\'{\i}fico e
Tecnol\'ogico (CNPq)}
\thanks{A. S. Bazanella is with the Department of Automation and Energy,
Universidade Federal do Rio Grande do Sul (DELAE/UFRGS), Porto Alegre-RS, Brazil. Email: bazanella@ufrgs.br}
\thanks{D. Eckhard is with the Department of Pure and Applied Mathematics,
Universidade Federal do Rio Grande do Sul (DELAE/UFRGS), Porto Alegre-RS, Brazil. Email: diegoeck@ufrgs.br}
}

\begin{document}

\maketitle
	\pagestyle{empty}
	\thispagestyle{empty}

%

\begin{abstract}                          
Virtual Reference Feedback Tuning (VRFT) is a  well known and very successful data-driven control 
design  method. It has been initially conceived for linear plants and this original formulation has been 
much explored in the literature, besides having already found many practical applications.
A nonlinear version of VRFT has been proposed early on, but not much explored later on.
In this paper we highlight various issues involved in the application of nonlinear VRFT and
propose the inclusion of $L_1$ regularization in its formulation.
We illustrate by means of two simple examples the critical role played by two aspects: 
the $L_1$ regularization and the choice of
dictionary used to describe the nonlinearity of the controller.  	
\end{abstract}

\section{Introduction}




Data-driven (DD) control is a research topic that has received considerable attention
at least since the early 1990's \cite{IFT_prime}, and that more recently has experienced
a large boost. Most of the early work on this subject turned around the classical model reference paradigm,
and many different design methods have been developed along this line, usually known by their
acronyms, like IFT \cite{IFT_prime}, CbT \cite{CbT}, VRFT \cite{VRFT_prime}, OCI \cite{franklin}, among others.
These methods have been extensively analyzed, enhanced, tested and applied, to the point that this has
become a rather mature field \cite{nosso_livro}. 
More recently, DD control theory started spreading towards various other control design paradigms, opening huge new 
perspectives and opportunities that are being intensely explored - see \cite{Dorfler2022, Frank2021, Mircea2022, Camlibel2021}, for example. 
Meanwhile, the more traditional model reference approach is still seeing novelties, and this paper follows along this line. 

	In this paper we concentrate on Virtual Reference Feedback Tuning (VRFT), which is arguably the simplest, most well-known, 
	and most widely applied data-driven control design method. VRFT has been first presented in \cite{VRFT_prime} 
	for linear plants. Many later publications have extended VRFT's scope and improved its performance: it has been extended for nonmnimum phase plants in
	\cite{VRFT_NMP} and for multivariable plants in \cite{MIMO_NMP, Savaresi_MIMO}, and reformulated to optimize the response to disturbances in \cite{VDFT}.  
	It has also been adapted to obtain improved statistical properties \cite{Cristiane_JCAES}. VRFT has been very successful in various practical applications using this linear
	formulation \cite{Savaresi_protheses, Savaresi_braking, Savaresi_engine}.
	
	A nonlinear version of VRFT appeared in \cite{VRFT_nonlinear}, but this nonlinear design method
	has not received similar attention in the literature. The application of the method in a nonlinear setting presents new challenges that, to the most part,
	have not yet been properly described, let alone solved. These challenges are more easily perceived recalling that VRFT,
	like all methods based on the model reference paradigm, can be seen as estimation of an {\em ideal controller}. The ideal controller is the one that, if put in the loop, would provide
	exactly the performance that has been specified. In the linear setting, the ideal controller is a linear object, so it is rather easy to devise a controller
	parametrization that would be appropriate to estimate it approximately. Moreover, this parametrization will be of very low dimension for the majority of
	problems - never larger than twice the order of the plant. It is also possible, though not so easy, to understand how undermodeling affects the closed-loop
	performance and how to cope with this issue \cite{nosso_livro}. 
	
	When the object to be estimated - in this case the ideal controller - is a nonlinear map, one has to be concerned not only with its dynamic order but also 
	with the nature of the nonlinearities involved. These are mostly unknown, since no knowledge of the plant's model can be assumed. 
	As a result, even the simplest examples will require large dictionaries to obtain a decent estimator, which has two major consequences.
	One is that the resulting controller, with such large number of parameters, is in most cases very undesirable and in many cases not viable. 
	The other one is that the statistical properties of the estimate will tend to be poor.
	The use of regularization is thus in order to obtain more appropriate statistical properties and more parsimonious controllers. 
	
	In this paper we explore the application of VRFT to nonlinear plants. We discuss the above issues of parametrization 
	and optimization, and propose the use of Lasso to solve the VRFT design problem.
	The paper is organized as follows. In Section \ref{sec:problem} we introduce the model reference control problem for a nonlinear plant.
	This sets the stage to present, in  \ref{sec:VRFT}, the VRFT method in a general nonlinear setting, with the inclusion of $L_1$ regularization,
	that we call the LASSO-VRFT. 
	In Section \ref{sec:cases} we illustrate the performance of VRFT and LASSO-VRFT in two simulation case studies.
	These are simple examples that serve to highlight
	the possibilities and limitations of each tool: VRFT and its LASSO counterpart, dictionaries, regularization. Some conclusions and future directions
	of work are briefly discussed in Section \ref{sec:conc}.

\section{Model reference control design}\label{sec:problem}

%

We are given a discrete-time SISO plant, which can be described as
\begin{equation}\label{plant}
y(t) = P[ u(t)] +\nu(t)
\end{equation}
where $y(t)$ is the output, which must track a reference signal $r(t)$, 
$u(t)$ is the control input, and $\nu(t)$ is the measurement noise, a stationary process. 
The input-output map $P[\cdot ]$ is  such that, for any finite control signal $u(t)$,
the solutions of (\ref{plant}) exist and are unique. 

The closed-loop performance is specified by the reference model $T_d(z)$, which
is a transfer function representing the desired input-output behavior in closed-loop, that is, 
from the reference $r(t)$ to the output $y(t)$.
The control objective is thus to make the closed-loop map from $r(t)$ to $y(t)$ to be
as close as possible to the given linear map $T_d(z)$, while maintaining internal stability.

It has been shown in \cite{VRFT_nonlinear} that, if the map $P[\cdot ]$ is invertible, there exists 
an {\em ideal controller}
$$
u(t) = C_0[r(t)-y(t) ]
$$
which results in the desired closed-loop behavior if put in the loop. Thus, 
the design of a controller in the model reference framework can be seen as the exercise of estimating 
this ideal controller. 

We define the control law, which is aimed at approximating the ideal controller, in a linearly parameterized way:
\begin{equation}\label{controller_class}
u(t) = \Sigma_{i=1}^{m} \rho_i \psi_i(z(t)) \bydef C(\rho, z(t)) ,
\end{equation}
where $z(t)$ is the set of measurements, 
$\psi_i(z(t)), i = 1, \ldots , m$ contains a dictionary of functions and/or 
functionals, and $\rho = [\rho_1 ~\ldots ~ \rho_m]^T$ is the parameter vector. 

Given a reference model $T_d(z)$, a controller class in the form (\ref{controller_class}),
and input-output data collected from the plant $u(t), y(t), ~ t=1, \ldots, N$, the best controller will
be the one  minimizing the cost function
\begin{equation}\label{MR_cost}
J(\rho ) =  \sum_{t=1}^N [y_d(t) - y(t, \rho) ]^2 
\end{equation}
where $y_d(t)= \mathcal{Z}^{-1} \{T_d(z)R(z)\}$ is the
desired closed-loop response and $y(t, \rho)$
is the actual closed-loop response to $r(t)$
obtained with the control law $C(\rho, e(t) )$.

\section{Nonlinear VRFT and Regularization}\label{sec:VRFT}

\subsection{The VRFT concept}

The cost function $J(\rho )$ in \eqref{MR_cost} is dependent, through $y(t,\rho )$, on the plant model, which is 
not available for the designer. Moreover, 
it is nonconvex, even with a linearly parameterized controller as (\ref{controller_class}).\footnote{This happens even if the plant is linear}
VRFT is a design method that substitutes the cost function $J(\rho )$ with the function
\begin{equation}\label{cost_VRFT}
J^V(\rho) = \sum_{t=1}^N [u(t)-(\Sigma_{i=1}^{m} \rho_i \psi_i(\bar{z}(t)) ]^2
\end{equation}
where $\bar{z}(t)$ is a virtual version of the measurement, that is, one in which
every instance of  the reference signal $r(t)$ is substituted by  the virtual reference
$\bar{r}(t) = \mathcal{Z}^{-1} \{T_d^{-1}(z)Y(z) \}$.
Under ideal conditions $J^V(\cdot )$ has the same global minimum as $J(\cdot )$,
but $J^V(\cdot )$ does not depend explicitly on the plant model, so it can be minimized without
knowledge of such model, using only input-output data from the plant.
Moreover, with a linear parametrization like (\ref{controller_class}), this is a quadratic
function of the parameters to be designed and thus can be minimized by least squares. 

The VRFT method is well-known and has been extensively applied to all sorts of linear plants, 
including MIMO and nonminimum-phase plants \cite{Savaresi_protheses, Savaresi_braking, Savaresi_engine}.
These applications include many experimental
ones and also actual industrial applications, so it is a well established and successful control design
methodology. For nonlinear plants and controllers, the theory has been 
provided in \cite{VRFT_nonlinear} but applications are hard to find in the literature  \cite{Tassiano}, even at the simulation
level.

%
%

\subsection{The VRFT Regression}


The VRFT control design consists in solving the regression defined by \eqref{cost_VRFT}.
In the nonlinear setting, a large dictionary of nonlinear functions must be used in most cases.
Indeed, without knowledge of a model for  the plant, one does not know a priori which nonlinear functions
must be present in the dictionary, so a large number of candidate functions must be employed.
This contrasts with the linear case, in which a ``linear dictionary'' is used and thus the number of terms
is at most twice the controller's order, which is usually quite low. 
Priors on the plant's nature will be very welcome to allow a reasonable choice of the dictionary's
structure - polynomial, trigonometrical, rational, etc. Without any priors, all sorts of functions must
be included. 

But even if one restricts the dictionary to a particular class of functions, the dimension
$m$ grows very rapidly.
Take the example of a first order error feedback controller; then there are typically three signals in $z(t)$:
$e(t)$, $e(t-1)$ and $u(t-1)$. If we take a modest third-order polynomial dictionary we'll have
nineteen terms already. With a second-order error feedback controller and the same third-order
polynomial structure for the dictionary, 251 terms.
Ordinary least squares is unlikely to handle properly such quantities of parameters, and
the statistical properties 
will likely deteriorate very rapidly. Moreover, even if least squares would provide a statistically sound solution,
one does not want to implement a controller with more than two hundred parameters; a more parsimonious
controller is desired in any case.
Thus, some sort of regularization is asked for, to achieve parsimony and improved statistical properties. 
\footnote{Though this issue is outside the scope of this paper, it is worth mentioning that regularization can also play
a more conceptual and versatile role in DD control design \cite{Dorfler2022}.}


\subsection{LASSO-VRFT}


In order to cope with the large dimension of the dictionary and the case in which the ideal controller
is not in the controller set, we propose to apply $L_1$  regularization to the optimization cost $J^V(\cdot )$.
The least squares regression with $L_1$ regularization is also known as the LASSO - acronym for 
Least Absolute Shrinkage and Selection Operator - and consists, in our case,  in
minimizing the cost function
\begin{equation}\label{cost_LASSO}
J^V_L(\rho) = \sum_{t=1}^N [u(t)-(\Sigma_{i=1}^{m} \rho_i \psi_i(\bar{z}(t)) ]^2 + \alpha \sum_{i=1}^m | \rho_i |
\end{equation}
where $\alpha$ is the shrinking factor, to be chosen. Any regularization reduces the magnitudes of the 
parameters. Due to its geometric features, the $L_1$ norm
tends to yield solutions in which some parameters are exactly zero. Thus, the LASSO is known
to work as a selection mechanism, providing parsimony to the solution \cite{stat_learn}. 

The larger the value of $\alpha$, the more parameters will be exactly zero. 
In this work, the Python package Scikit was used to minimize the criterion \ref{cost_LASSO}.
This package uses the coordinate descent algorithm to find the parameters that minimize the LASSO criterion.
All simulations in this work used $\alpha=0.001$, which allowed to reduce the number
of parameters without a significant loss of performance, and the optimization algorithm was
limited to use a maximum of $100,000$ iterations.
Some techniques for automatic choice of $\alpha$ are still under investigation,
including the Akaike information criterion (AIC), the Bayes Information criterion (BIC) and Cross Validation techniques.

\section{Case studies}\label{sec:cases}

\begin{figure}[!h]
    \centering
    \includegraphics[width=1\columnwidth]{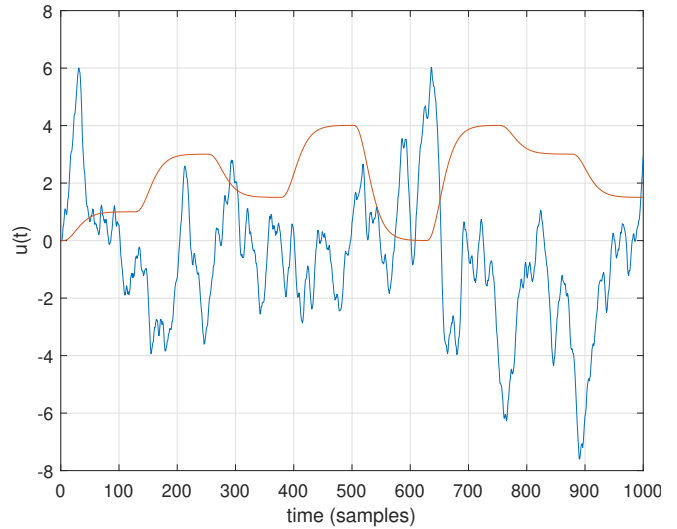}
    \caption{The two different inputs applied at both plants: filtered steps (dashed line) and filtered white noise (continuous line).}
    \label{fig:all_inputs}
\end{figure}

In this Section we present simulation studies for two different plants. 
We collect input-output data from two discrete-time nonlinear plants excited by two different input signals:
a uniformly distributed zero-mean random sequence, and a sequence of steps, both 
filtered by a transfer function $$F(z) = T_d(z) (1-T_d(z)) \frac{a}{z-1} ,$$ as recommended
in \cite{VRFT_nonlinear}, and for numerical reasons we tune the parameter $a$
so that $F(1)=1$. These two input signals are presented in Figure \ref{fig:all_inputs}, and they are both 
of length $N=1,000$. The output measurement is contaminated by zero-mean gaussian white noise with 
varying energy levels. Each one of the following subsections presents the results for a different plant.
The results obtained with the two input signals are very similar, so we will present
in detail only the ones obtained with the random input.

For both plants there is a specification of zero steady state tracking error for constant references, which
requires an integrator in the controller. We thus define the simplest controller structure possible under this constraint, 
which is an integrator followed by a nonlinear static element, that is, the measurement set is just
$z(t) = \sum_{\tau =1}^{t-1} e(\tau )$, the control law is $u(t) = \varphi( z(t))$, and the nonlinear function
is what must be estimated by the VRFT regression: $\varphi( z(t)) \approx  \sum_{i=1}^m \rho_i\psi_i(z(t))$.

The exact plant models are not used in any design,
so they are not relevant to our presentation, except for transparency; their presentation is thus delayed to Subsection
\ref{sec:examine}. We also  collect data from both plants with a (nonfitered) step sequence just to have a ballpark estimate of the settling time. 
The step responses  reveal settling
times around twenty-five samples for both plants, with plant \#1 faster than plant \#2,
an information that will determine our choice of reference model. 

\paragraph{The reference model}
Since the closed-loop system must track constant references, the reference model must satisfy $T_d(1)=1$.
The reference model must have the same relative degree of the open loop system formed by 
plant and controller \cite{Gustavo_TCST}.
The relative degree of the controller is one and any sampled system of finite order has relative degree
one. Accordingly, we assume that the plants have no delay and specify a reference model with relative
degree equal to two. Since we have no other information about the plant and the controller is the simplest possible,
a conservative transient specification is in order. So we pick a pole at $z=0.9$, which corresponds to a settling time
of 37 samples, considerably larger than the settling time in open-loop. As a result of these considerations, the reference model is chosen as
$$
T_d(z) = \frac{0.01}{(z-0.9)^2} 
$$
for both plants.

\paragraph{The dictionary}
Next, we need to choose the dictionary of functions $\psi_i(\cdot )$. We start with a polynomial
dictionary, which is quite general and intuitive, as polynomials form an orthogonal basis for the
space of analytical functions. We observe that the data are symmetrical around zero,
which suggests that the plant's nonlinearity has odd symmetry. This implies that
the optimal $\varphi(\cdot )$ also has odd symmetry, and thus only odd powers are needed
in the dictionary. For numerical  reasons, we normalize the basis such that all 
elements have unitary magnitude at the end of scale for the data that have been collected. 
From all these considerations 
we arrive at the following parametrization:
\begin{equation}\label{poly_dic}
\hat{\varphi}(z) = \sum_{i=1}^m \rho_i (\frac{z}{200})^{2i-1} 
\end{equation}

\subsection{Ideal scenario - plant \#1}

We have run the standard VRFT  regression \eqref{cost_VRFT}  
and the LASSO-VRFT \eqref{cost_LASSO} with the
polynomial dictionary using various values of $m$, using the data originated from the 
random input. 
For each design we assess the performance of each controller by evaluating 
the closed-loop response to a sequence of reference steps. 

\begin{figure}[!h]
    \centering
     \includegraphics[width=1\columnwidth]{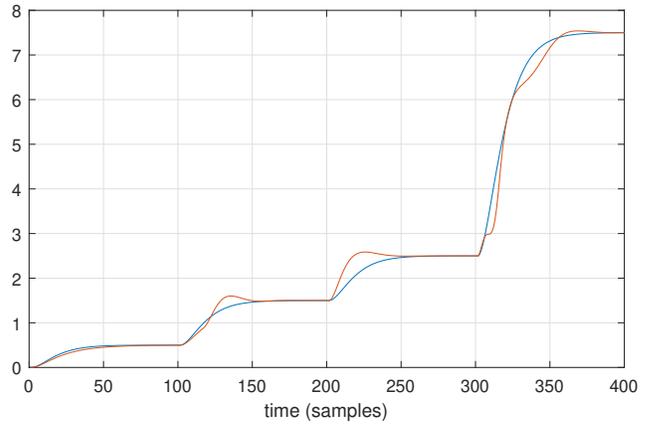}
    \caption{Closed-loop response of plant \#1  with the controller obtained from the polynomial dictionary with $m=20$}
    \label{fig:data1stepspoly}
\end{figure}

The closed-loop response resulting from the controller obtained with $m=20$ is presented in Figure \ref{fig:data1stepspoly},
along with the desired response $y_d(t)={\cal Z}^{-1}\{T_d(z)R(z)\}$; the performance is not impressive.
This somewhat expensive controller still results in a large
error with respect to the desired response, and we have found that  increasing $m$ improves the performance very slowly. 
In all cases, when comparing the results with VRFT and with the LASSO-VRFT, we have observed that the regularization did not reduce
significantly the number of parameters. So, we have been left with either a poor performance or with hundreds of parameters - hence
no satisfactory solution.  We have repeated the experiments with the filtered step input and they have yielded very similar results. 
We have also repeated these experiments with various levels of output noise, as measured by the noise's variance $\sigma$, with 
similar results up to $\sigma = 0.1$. For larger noise energies the closed-loop performance deteriorates rapidly in all designs.
From now on all results presented are for $\sigma=0.05$.

\begin{figure}[!h]
    \centering
    \includegraphics[width=1\columnwidth]{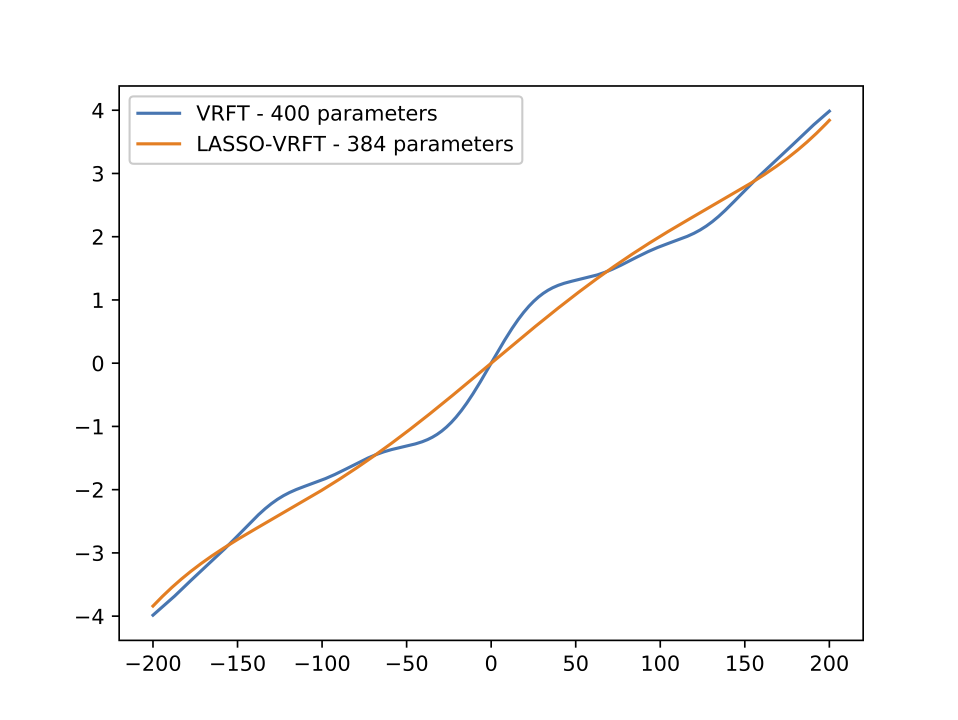}
    \caption{Estimated map $\hat{\varphi}(\cdot )$ by the polynomial dictionary 
    for plant \# 1 with $m=400$: VRFT (blue) and LASSO-VRFT (orange)}
    \label{fig:phihat400poly}
\end{figure}

Judging by the observation of the data, this is not a plant with a complex dynamics. Yet, even if we collect data 
free of noise we are not able to achieve a parsimonious solution with good performance. 
Clearly, the controller structure is very simple, which limits severely the performance that can be achieved, but in fact 
the choice of dictionary is also to blame for this disappointing result.
In order to assess the dictionary, we plot the estimated function $\hat{\varphi} (\cdot )$ obtained with $m=400$,
seen in Figure \ref{fig:phihat400poly}. The regularization in the LASSO-VRFT has only mildly reduced the number of parameters,
to $384$, and yet it can be seen in the Figure that it has changed considerably the character of the estimated function.  
Thus, it can be inferred that the $384$ terms are necessary to provide a good approximation for the 
function $\varphi(\cdot )$.

%
%

\begin{figure}[!h]
    \centering
    \includegraphics[width=1\columnwidth]{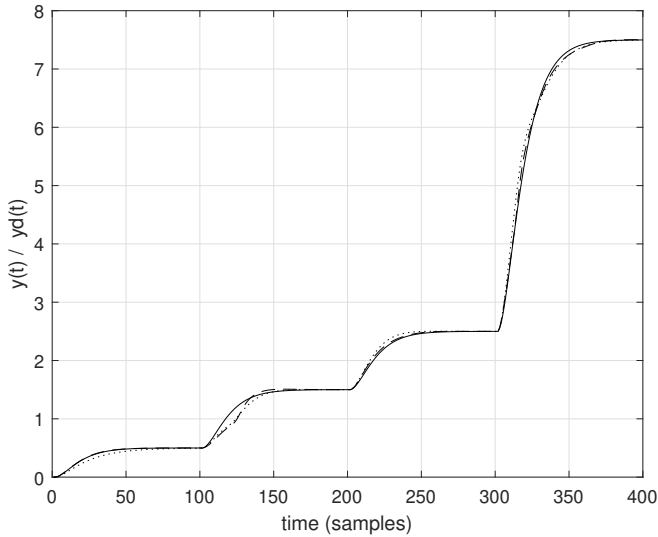}
    \caption{Closed-loop responses of plant \#1 with the controllers obtained from the deadzone dictionary with $m=20$, with
    VRFT (dashed line) and LASSO-VRFT (dotted line)}
    \label{fig:data1stepsZM20}
\end{figure}

This plot looks like a piecewise affine function, which is hard to approximate  with a polynomial basis, since it is not analytic. 
We thus pick a different dictionary, more suited to describe piecewise affine functions, which 
is formed by various deadzones. This dictionary is 
\begin{equation}\label{ZM_dic}
\hat{\varphi}(x ) =  \sum_{i=1}^{m} \rho_i ~ ZM_i(x )
\end{equation}
where $ZM_i(\cdot )$ is the deadzone nonlinearity:
\begin{equation}\label{base_ZM}
ZM_i( x) =  \left\{\begin{array}{ll} \frac{x+10 (i-1)}{200-10(i-1)} & x < -10 (i-1) \\ 
						\frac{x-10 (i-1)}{200-10(i-1)} & x >10 (i-1) \\
 						0 & -10 (i-1) < x < 10 (i-1)  \end{array} \right. .
\end{equation}
and we choose initially $m=20$.
We run VRFT and LASSO-VRFT again with the same data and $\sigma=0.05$, now with this new dictionary. 
The responses of the closed-loop system with the resulting controllers,
to the same reference steps as those in Figure \ref{fig:data1stepspoly},
are presented in Figure \ref{fig:data1stepsZM20}.
The performance is very close to the desired one and
the regularization has reduced the number of parameters from twenty to only four 
without significant change of performance.

\begin{figure}[!h]
    \centering
        \includegraphics[width=1\columnwidth]{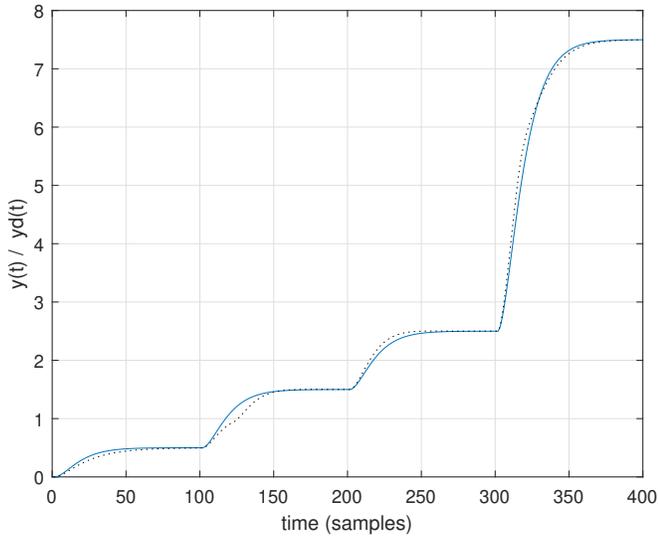}
    \caption{Closed-loop response of plant \#1 with the controller obtained from the deadzone dictionary with $m=400$ with LASSO-VRFT}
    \label{fig:data1stepsZM400}
\end{figure}

With this we have successfully concluded a data-driven design with LASSO-VRFT. 
In order to learn more about the behavior of the dictionary and the role of the regularization, we have tried a larger dictionary 
with the deadzone basis: $m=400$. This new dictionary will have the same definition as is \eqref{ZM_dic} but with each occurrence
of the number $10$ replaced by $0.5$. We have run the LASSO-VRFT (VRFT is not of interest for this exercise) and obtained the
results in Figure \ref{fig:data1stepsZM400}, with the regularization having reduced
the number of nonzero parameters to 47. The regularization was successful in reducing drastically the
number of nonzero parameters and providing good closed-loop performance.
Equally important is the fact that the regularization has proven  indispensable in this case:
the controller obtained without regularization presented many parameters
with very large magnitudes and resulted in an unstable closed-loop.

\subsection{Non-ideal scenario - plant \#2}

\begin{figure}[!h]
    \centering
    \includegraphics[width=1\columnwidth]{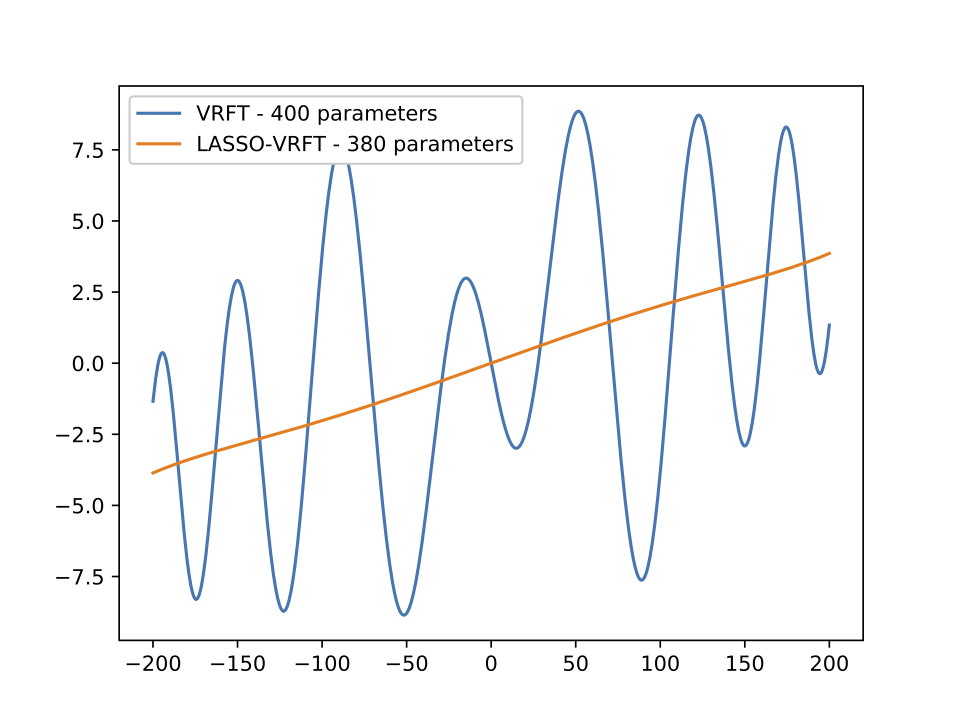}
    \caption{Estimated map $\hat{\varphi}(\cdot )$ with polynomial dictionary $m=400$ for plant \# 2: VRFT and LASSO-VRFT}
    \label{fig:varphi2}
\end{figure}

When we run LASSO-VRFT with the polynomial dictionary, the map's estimate $\hat{\varphi}(\cdot )$ shown
in Figure \ref{fig:varphi2} is obtained. Notice how the regularization zeroes only twenty
parameters and how this completely changes the map's character to an almost linear 
map. Moreover, the controller obtained without regularization is not even stabilizing.
In this case the polynomial dictionary is a complete failure: even if we were willing
to implement a controller with several hundred parameters, the result would either
be an unstable closed-loop (VRFT) or poor performance with no compensation
of the nonlinear features of the plant (LASSO-VRFT).

\begin{figure}[!h]
    \centering
    \includegraphics[width=1\columnwidth]{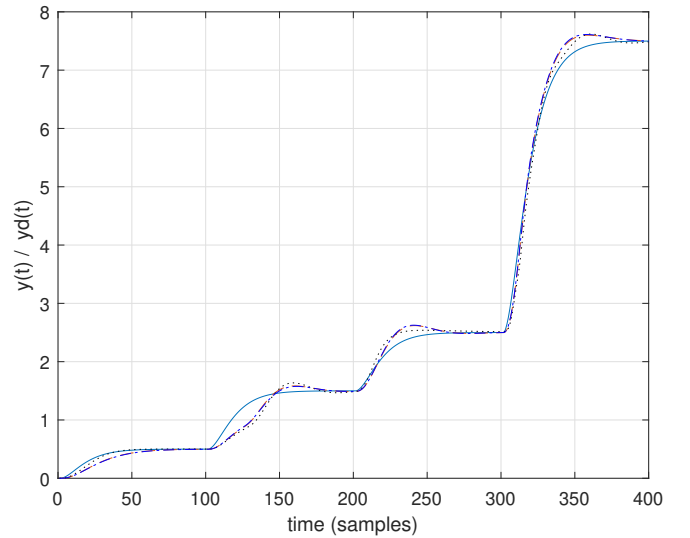}
    \caption{Closed-loop responses of plant \#2 with the controllers obtained from the deadzone dictionary: $m=20$ with VRFT (dotted line) and
    with LASSO-VRFT (dashed);  $m=400$, with LASSO-VRFT (dash-dot)}
    \label{fig:data2stepsZM}
\end{figure}

We next try the deadzone dictionary \eqref{ZM_dic}, first  with $m=20$ and then with $m=400$.
The LASSO-VRFT reduced the number of parameters from $400$ to  $52$, and from $m=20$ to $8$.
The closed-loop performance is shown in Figure \ref{fig:data2stepsZM}, where it is seen that 
the performances obtained with the LASSO-VRFT with $m=20$ and $m=400$ are virtually 
indistinguishable.

\subsection{Examination}\label{sec:examine}

Though the plants' models were used to get the data, the control design was performed without using any knowledge of them, except
the assumption that the plants were delay-free. Let us now disclose what are the plants and the ideal controllers to better understand what happened and why.

Both plants consist of Hammerstein systems, with the same piecewise affine nonlinear function 
$$
\phi(x) = \left\{\begin{array}{ll} 	2 x - 2 ,& x < -2  \\
						5 x + 4 ,& -2 < x < -1 \\
						x ,& |x| < 1 \\
						5 x - 4 ,& 1 < x < 2 \\
						2 x + 2 ,& x > 2 \end{array}\right.
$$
in front of a linear block. 

The linear block of plant \#1 is a first-order system with transfer function
$$
G(z) = \frac{0.2}{z-0.8} 
$$
whereas in plant \#2  the transfer function is
$$
G(z) = \frac{0.04~z}{(z-0.8)^2} .
$$

It can be verified without much effort that the ideal controller is in the controller class for plant \#1, and is given by:
\begin{eqnarray}
v(t) & = & v(t-1) + 0.05 e(t) \\
u(t) &=&  \phi^{-1} (v(t))
\end{eqnarray}
where $\phi^{-1}(\cdot )$ is the left-inverse of $\phi(\cdot )$, that is, $\phi^{-1}(\phi(x))=x$. 
It is easily seen that the nonlinear function presented in Figure \ref{fig:phihat400poly} is a close approximation 
to the left inverse of $\phi (\cdot )$ multiplied by a constant factor $0.05$. Hence, the polynomial dictionary 
estimated a controller that is very close to the ideal controller, but for that it needed hundreds of parameters.
On the other hand, the deadzone dictionary was able to get very close to the ideal controller with a parsimonious controller; 
this was true also for the case $m=400$.
The choice of dictionary has played a critical role for this very simple example, and the $L_1$ regularization was
also very important.

Concerning plant \#2, the ideal controller does not belong to the controller class; it is given by
\begin{eqnarray}
V(z) & = & \frac{0.25 (z-0.8)}{z(z-1)} E(z)\\
u(t) &=&  \phi^{-1} (v(t))
\end{eqnarray}
where $V(z) = {\cal Z}\{v(t)\}$ and $E(z) = {\cal Z}\{e(t)\}$.
For this plant, regularization played a more critical role than for the first plant, where the ideal controller belongs to the controller class. 
Indeed, without regularization both dictionaries were unable to provide a stabilizing controller. 

\section{Conclusions}\label{sec:conc}

We have discussed the difficulties found in the application of the VRFT method
to nonlinear plants, and proposed the inclusion of $L_1$ regularization in the regression problem
to cope with these difficulties. Two simple case studies have been presented that illustrated
these difficulties and highlighted two main points: the important role played by the regularization and the
criticality of the choice of dictionary to describe the controller.
For the case in which the ideal controller is not in the control class, regularization proved 
to be more critical to the success of the design. Future work concentrates on the 
application of these ideas to the design of controllers for more complex plants, also
with other DD control design methods. 

\bibliographystyle{plain}        
\bibliography{../../bibtex/databased, ../../bibtex/ID}

\end{document}